Computational Logic 3:

# SAT Techniques for Lexicographic Path Orders

Seminar Report

Harald Zankl

29th December 2017

**Supervisor:** Univ.-Prof. Dr. Aart Middeldorp

## Abstract

This report was written for the seminar `Computational Logic 3` conducted at the Institute of Computer Science of the University of Innsbruck. It is concerned with a SAT solver approach for deciding LPO-termination of term rewrite systems and a continuation of the seminar report [13] which considers a BDD approach. After relevant algorithms are explained, experimental results are reported.

# Contents

# 1 Introduction

Termination is one of the most important properties of term rewrite systems (TRSs). In general it is undecidable but for certain classes of TRSs powerful methods have been developed to decide termination. One of these methods is the lexicographic path ordering (LPO). In this seminar report an alternative algorithm to the one given in [6] for deciding LPO-termination suggested by Codish *et al.* in [2] is discussed. Implementations of both encodings are compared to the standard T<sub>T</sub>T implementation ([10]) on a database of 773 TRS instances. The main idea of the approaches in [2] and [6] is to extract the constraints LPO puts on a precedence for a given TRS into a propositional formula. Afterwards some more constraints are added to ensure the properties of a precedence and finally the propositional formula is tested for satisfiability. The two approaches differ in the way the additional constraints are expressed and how the result is tested for satisfiability. [6] uses binary decision diagrams (BDDs) whereas [2] integrates the SAT solver MiniSat [3].

In Section 2 some simple definitions are fixed and results mentioned. Section 3 describes how to get the constraints for LPO-termination and explains how the additional constraints are constructed. After a short discussion about optimisations in Section 4, the run time results are presented in Section 5. Some remarks about the paper [2] can be found in Section 6. The report is concluded with ideas for future work which are mentioned in Section 7. Appendix A contains a proof that the version of LPO defined here is indeed a simplification order and Appendix B shows how C++ can be interfaced with `OCaml`.

# 2 Preliminaries

## 2.1 Relations

**Definition 1.** A *quasi-order* is a reflexive and transitive relation. A *proper order* is an irreflexive and transitive relation. An *equivalence relation* is reflexive, symmetric, and transitive.

**Lemma 1.** *Let $\succsim$ be a quasi-order on a set $A$. Then $\{a \succ b \mid a \succsim b$ and not $b \succsim a\}$ is a proper order and $\{a \sim b \mid a \succsim b$ and $b \succsim a\}$ is an equivalence relation on $A$. Furthermore $\succsim \,=\, \succ \uplus \sim$ where $\uplus$ denotes disjoint union.* □

**Remark 1.** When talking about quasi-orders it sometimes is convenient to specify the strict and equivalence part separately. So $\mathsf{f} \succsim \mathsf{f}$, $\mathsf{g} \succsim \mathsf{g}$, $\mathsf{h} \succsim \mathsf{h}$, $\mathsf{f} \succsim \mathsf{g}$, $\mathsf{g} \succsim \mathsf{f}$, $\mathsf{f} \succsim \mathsf{h}$, and $\mathsf{g} \succsim \mathsf{h}$ is then written as $\mathsf{f} \sim \mathsf{g}$, $\mathsf{f} \succ \mathsf{h}$, and $\mathsf{g} \succ \mathsf{h}$ or even as $\mathsf{f} \sim \mathsf{g}$ and $\mathsf{f} \succ \mathsf{h}$ because $\mathsf{g} \succ \mathsf{h}$ follows implicitly from the transitivity of $\succsim$.

Next the lexicographic extension of a quasi-order is defined.

**Definition 2.** Let $\succsim$ be a quasi-order. The *lexicographic extension* $\succsim^{\mathsf{lex}}$ is then defined as follows:

$$\langle s_1, \ldots, s_m \rangle \succsim^{\mathsf{lex}} \langle t_1, \ldots, t_n \rangle \iff m > 0 \wedge$$
$$\Big(n = 0 \vee \Big(n > 0 \wedge \Big(s_1 \succ t_1 \vee \Big(s_1 \sim t_1 \wedge \langle s_2, \ldots, s_m \rangle \succsim^{\mathsf{lex}} \langle t_2, \ldots, t_n \rangle\Big)\Big)\Big)\Big)$$

When dealing with proper orders, equivalence amounts to equality.

## 2.2 LPO

Let $\mathcal{F}$ be a signature and $\mathcal{V}$ a set of variables. Then $\mathcal{T}(\mathcal{F}, \mathcal{V})$ denotes all terms which can be build over $\mathcal{F}$ and $\mathcal{V}$. A *term rewrite system* (TRS) $\mathcal{R} = (\mathcal{F}, R)$ consists of a signature $\mathcal{F}$ and a set of rewrite rules $R \subseteq \mathcal{T}(\mathcal{F}, \mathcal{V}) \times \mathcal{T}(\mathcal{F}, \mathcal{V})$. We write $l \to r \in R$ instead of $(l, r) \in R$.
The basic idea of term rewriting is to apply these rules to terms. We skip the details of how they are applied. A general question which arises is whether this process of applying rules stops at some time. We call this (undecidable) property of $\mathcal{R}$ *termination.* For several classes of TRSs powerful methods have been developed to determine if the given instance is terminating. In the sequel we focus on the lexicographic path ordering which provides a sufficient condition for termination. (As it is not a necessary condition there are terminating TRSs which cannot be proved terminating by LPO.)

**Definition 3.** A *quasi-precedence* $\succsim$ (*strict precedence* $\succ$) is a quasi-order (proper order) on a signature $\mathcal{F}$. Sometimes we find it convenient to call a quasi-precedence simply precedence.

Note that as an immediate consequence of Lemma 1 we can build a quasi-precedence out of a strict one by adding reflexivity. Therefore all results carry over. Next the induced order $\succsim_{\text{lpo}}$ for a given precedence will be defined. Therefore we split it into its strict ($\succ_{\text{lpo}}$) and its equivalence part ($\sim_{\text{lpo}}$). First consider $\sim_{\text{lpo}}$.

**Definition 4.** Let $\succsim$ be a precedence and $s, t \in \mathcal{T}(\mathcal{F}, \mathcal{V})$. We define $s \sim_{\text{lpo}} t$ if one of the following alternatives holds:

(1) $s = t$, or

(2) $s = f(s_1, \ldots, s_m), t = g(t_1, \ldots, t_m), f \sim g$, and $s_i \sim_{\text{lpo}} t_i$ for all $1 \leq i \leq m$.

After that the strict part can be defined as follows:

**Definition 5.** Let $\succsim$ be a precedence and $s, t \in \mathcal{T}(\mathcal{F}, \mathcal{V})$. If $s,\ t \notin \mathcal{V}$ then $s = f(s_1, \ldots, s_m)$ and $t = g(t_1, \ldots, t_n)$. We have $s \succ_{\text{lpo}} t$ if one of the following alternatives holds:

(1) $f \sim g$ and either there is an $i \in \{1, \ldots, m\}$ such that $s_j \sim_{\text{lpo}} t_j$ for all $1 \leq j < i$, $s_i \succ_{\text{lpo}} t_i$, and $s \succ_{\text{lpo}} t_j$ for all $i < j \leq n$, or $m > n$ and $s_i \sim_{\text{lpo}} t_i$ for all $1 \leq i \leq n$, or

(2) $f \succ g$ and $s \succ_{\text{lpo}} t_j$ for all $1 \leq j \leq n$, or

(3) there is an $i \in \{1, \ldots, m\}$ such that $s_i \succsim_{\text{lpo}} t$.

When $s \succ_{\text{lpo}} t$ by clause $(j)$ in the definition above, we write $s \succ_{\text{lpo}} t \ \langle j \rangle$. If convenient we add the index $i \in \{1, \ldots, m\}$ in cases (1) and (3). Be aware that $\langle 1 \rangle$ might refer to case (1) where the second alternative applies or in a more general view where the exact alternative is not important. The context clarifies the exact meaning. When considering strict precedences $\succ$ the equivalence relation $\sim_{\text{lpo}}$ amounts to syntactical identity.

**Definition 6.** A TRS $\mathcal{R} = (\mathcal{F}, R)$ is called *(quasi-)LPO-terminating* if and only if there exists a precedence $\succsim$ such that $l \succ_{\text{lpo}} r$ for all $l \to r \in R$. In case that $\succsim$ is a proper order we speak of *strict LPO-termination.*

Note that as soon as the precedence $\succsim$ is fixed the induced order $\succsim_{\text{lpo}}$ is determined uniquely. In order to show that if a TRS is LPO-terminating it is actually terminating it suffices to show that $\succ_{\text{lpo}}$ is a simplification order. The proof can be found in Appendix A.

**Example 1.** Let $\mathcal{R}_1$ be the TRS consisting of the following rule:

$$\mathsf{f}(y, \mathsf{g}(x), x) \to \mathsf{f}(y, x, \mathsf{g}(\mathsf{g}(x)))$$

We want to determine if $\mathsf{f}(y, \mathsf{g}(x), x) \succ_{\text{lpo}} \mathsf{f}(y, x, \mathsf{g}(\mathsf{g}(x)))$. Clearly case (1) of the definition applies and the $i$ is 2. So it remains to test that $\mathsf{g}(x) \succ_{\text{lpo}} x$ and $\mathsf{f}(y, \mathsf{g}(x), x) \succ_{\text{lpo}} \mathsf{g}(\mathsf{g}(x))$. The first holds by (3). For the latter we use (2) and therefore we need $\mathsf{f} \succ \mathsf{g}$ in the (strict) precedence and $\mathsf{f}(y, \mathsf{g}(x), x) \succ_{\text{lpo}} \mathsf{g}(x)$ which again holds by (3). Hence this instance is strict LPO-terminating with strict precedence $\mathsf{f} \succ \mathsf{g}$ and LPO-terminating with quasi-precedence $\mathsf{f} \succsim \mathsf{f}$, $\mathsf{g} \succsim \mathsf{g}$, and $\mathsf{f} \succsim \mathsf{g}$.

Let $\mathcal{R}_2$ be the TRS consisting of the following two rules:

$$\mathsf{f}(x) \to \mathsf{g}(x)$$
$$\mathsf{g}(x) \to \mathsf{f}(x)$$

Strict LPO-termination is not the case since the first rule can only be handled by setting $\mathsf{f} \succ \mathsf{g}$ whereas the second one requires $\mathsf{g} \succ \mathsf{f}$. But as a strict precedence $\succ$ must be transitive, $\mathsf{f} \succ \mathsf{g}$ and $\mathsf{g} \succ \mathsf{f}$ yield $\mathsf{f} \succ \mathsf{f}$ which clearly contradicts irreflexivity. The system is also not quasi-LPO-terminating because if $\mathsf{f} \sim \mathsf{g}$ then the first rule does not fulfil $\mathsf{f}(x) \succ_{\text{lpo}} \mathsf{g}(x)$. Anyway, it would be strange if any formalism would state that the system is terminating because it obviously is not!

Let $\mathcal{R}_3$ be the TRS (from [2]) consisting of the following three rules:

$$\mathsf{div}(x, \mathsf{e}) \to \mathsf{i}(x)$$
$$\mathsf{i}(\mathsf{div}(x, y)) \to \mathsf{div}(y, x)$$
$$\mathsf{div}(\mathsf{div}(x, y), z) \to \mathsf{div}(y, \mathsf{div}(\mathsf{i}(x), z))$$

Again the system is not strict LPO-terminating because the first rule demands $\mathsf{div} \succ \mathsf{i}$ and the second one $\mathsf{i} \succ \mathsf{div}$. But it is LPO-terminating with quasi-precedence $\mathsf{div} \sim \mathsf{i}$.

# 3 Algorithms for LPO Termination

In this section at first an LPO encoding is presented in order to extract the demands on a precedence. Then the approach of [2] is reviewed to ensure the necessary properties of a precedence. In this approach function symbols are interpreted as natural numbers and then ordered by the greater or equal relation. Thus reflexivity and transitivity are enforced automatically.

## 3.1 LPO Encoding

As we have seen in the previous section we need to represent a precedence in order to decide LPO-termination. The following definition will take care of encoding a relation in propositional logic.

**Definition 7.** Let $X = \{X_{fg} \mid f, g \in \mathcal{F} \text{ with } f \neq g\}$ be a set of *propositional variables* and $Y$ such a similar set. An assignment $\alpha$ induces relations on $\mathcal{F}$ as follows: $f \succ g$ if and only if $\alpha(X_{fg}) = \mathsf{T}$ and $f \sim g$ if and only if $\alpha(Y_{fg}) = \mathsf{T}$.

Equivalence of terms using the $Y_{fg}$ variables can then be encoded as follows:

**Definition 8.** Let $s, t \in \mathcal{T}(\mathcal{F}, \mathcal{V})$. If $s, t \notin \mathcal{V}$ then $s = f(s_1, \ldots, s_m)$ and $t = g(t_1, \ldots, t_n)$:

$$E(Y)_{s,t} = \begin{cases} \top & \text{if } s = t \\ Y_{fg} \wedge E(Y)_{s_1,t_1} \wedge \cdots \wedge E(Y)_{s_m,t_m} & \text{if } m = n \\ \bot & \text{otherwise} \end{cases}$$

Next we generalise the definition of the lexicographic order to terms. Note that there is a nested recursion between $C(X,Y)_{s,t}$ which encodes LPO and $LEX(X,Y)_{\langle s_1,\ldots,s_m\rangle,\langle t_1,\ldots,t_n\rangle}$ which takes care of comparing the arguments lexicographically.

**Definition 9.**

$$LEX(X,Y)_{\langle s_1,\ldots,s_m\rangle,\langle t_1,\ldots,t_n\rangle} = \begin{cases} \bot & m = 0 \\ \top & m > 0 \text{ and } n = 0 \\ C(X,Y)_{s_1,t_1} \vee & \\ \big(E(Y)_{s_1,t_1} \wedge LEX(X,Y)_{\langle s_2,\ldots,s_m\rangle,\langle t_2,\ldots,t_n\rangle}\big) & m > 0 \text{ and } n > 0 \end{cases}$$

Now all preparation for the LPO encoding is done and $C(X,Y)_{s,t}$ in the following definition is a propositional formula which exactly mirrors the constraints LPO puts on the precedence in order to ensure $s \succ_{\text{lpo}} t$. The definition below is as general as possible and can also cope with quasi-precedences. The slight differences for strict precedences are indicated in the boxes. The first line refers to strict precedences (note that no $Y$ variables are needed) and the second line to quasi-precedences. Consider the third branch of Definition 10 first. The two function symbols of the terms $s$ and $t$ are different. For strict precedences we demand $f \succ g$ and we skip the first $k-1$ equal arguments ($s \succ_{\text{lpo}} t_l$ for all $1 \leq l < k$ holds automatically if $s_l = t_l$ for all $1 \leq l < k$) and demand that $s \succ_{\text{lpo}} t_l$ for all $k \leq l \leq n$. In case of a quasi-precedence we do not know if $f \succ g$ or $f \sim g$. In the former case we do the same as for the strict case whereas in the latter we must find a strict decrease in the arguments. Because the first $k-1$ ones are equal they are trivially equivalent and do not yield a strict decrease. It might happen that the $k$-th arguments are equivalent and the strict decrease is in some later arguments. That is why the lexicographic extension is applied here. It ensures that the arguments with the strict decrease are found. Finally $s \succ_{\text{lpo}} t_l$ for all $k \leq l \leq n$ ensures that all remaining arguments of $t$ are smaller than $s$. The fourth branch of the definition considers equal function symbols which are trivially equivalent. For strict precedences the $k$-th argument must be the one with the strict decrease whereas for quasi-precedences it again might happen that $s_k \sim_{\text{lpo}} t_k$ and so on. Note that the constraints for each branch in the definition below make the nondeterministic definition of LPO (Definition 5) deterministic.

**Definition 10.** Let $s, t \in \mathcal{T}(\mathcal{F}, \mathcal{V})$. If $s, t \notin \mathcal{V}$ then $s = f(s_1, \ldots, s_m)$ and $t = g(t_1, \ldots, t_n)$:

$$C(X,Y)_{s,t} = \begin{cases} \bot & \text{if } s = t \text{ or } s \in \mathcal{V} \text{ or both } t \in \mathcal{V} \text{ and } t \notin \mathcal{V}ar(s) \\[1em] \top & \text{if } s \notin \mathcal{V},\ t \in \mathcal{V}ar(s) \\[1em] \multicolumn{2}{l}{CE(X,Y)_{s,t}} \\ \multicolumn{2}{l}{\vee \left( X_{fg} \vee \boxed{\begin{array}{c} \bot \\ Y_{fg} \wedge LEX(X,Y)_{\langle s_k,\ldots,s_m\rangle,\langle t_k,\ldots,t_n\rangle} \end{array}} \right) \wedge C_k(X,Y)_{s,t}} \\ & \text{if } s \neq t, s \notin \mathcal{V}, t \notin \mathcal{V}, \text{ and } f \neq g \\[1em] \multicolumn{2}{l}{CE(X,Y)_{s,t}} \\ \multicolumn{2}{l}{\vee \left( \boxed{\begin{array}{c} C(X)_{s_k,t_k} \\ LEX(X,Y)_{\langle s_k,\ldots,s_m\rangle,\langle t_k,\ldots,t_n\rangle} \end{array}} \wedge C_{k+1}(X,Y)_{s,t} \right)} \\ & \text{if } s \neq t, s \notin \mathcal{V}, t \notin \mathcal{V}, \text{ and } f = g \end{cases}$$

with

$$CE(X,Y)_{s,t} = \bigvee_{i=1}^{m} E(Y)_{s_i,t} \vee \bigvee_{i=1}^{m} C(X,Y)_{s_i,t}$$

and

$$C_l(X,Y)_{s,t} = \bigwedge_{j=l}^{n} C(X,Y)_{s,t_j}$$

where $k$ is the minimum value of $i$ ($1 \leq i \leq m$) with $s_i \neq t_i$.

According to Definition 6 it is sufficient to test $l \succ_{\text{lpo}} r$ for all rules $l \to r \in R$ to ensure LPO-termination. The next definition expresses the quasi-LPO as well as the strict LPO constraints not only for single rules but for whole TRSs (setting $\emptyset$ for $Y$ suggests that no $Y$ variables are used, i.e., they all evaluate to $\mathsf{F}$).

**Definition 11.**

$$C(X,Y) = \bigwedge\{C(X,Y)_{l,r} \mid l \to r \in R\} \text{ and } C(X) = C(X,\emptyset)$$

**Example 2.** Let's first focus on strict LPO-termination. Therefore we compute $C(X,\emptyset) = C(X)$ for the first two TRSs of Example 1 given in the previous section. Note that as we are concerned with strict precedences $E(Y)_{s,t}$ is abbreviated to $E_{s,t}$ because no $Y$ variables are allowed. Consequently, two terms can only be equivalent if they are equal. Using the abbreviations $s = \mathsf{f}(y,\mathsf{g}(x),x)$ and $t = \mathsf{f}(y,x,\mathsf{g}(\mathsf{g}(x)))$ we get for $\mathcal{R}_1$

$$\begin{aligned}
C(X) &= C(X)_{s,t} \\
&= CE(X)_{s,t} \vee \big(C(X)_{\mathsf{g}(x),x} \wedge C_3(X)_{s,t}\big) \\
&= E_{y,t} \vee E_{\mathsf{g}(x),t} \vee E_{x,t} \vee C(X)_{y,t} \vee C(X)_{\mathsf{g}(x),t} \vee C(X)_{x,t} \\
&\quad \vee \big(\top \wedge C(X)_{s,\mathsf{g}(\mathsf{g}(x))}\big) \\
&= \bot \vee \bot \vee \bot \vee \bot \vee \big(CE(X)_{\mathsf{g}(x),t} \vee X_{\mathsf{gf}} \wedge C_1(X)_{\mathsf{g}(x),t}\big) \vee \bot \\
&\quad \vee C(X)_{s,\mathsf{g}(\mathsf{g}(x))} \\
&= \big(E_{x,t} \vee C(X)_{x,t} \vee X_{\mathsf{gf}} \wedge C(X)_{\mathsf{g}(x),y} \wedge C(X)_{\mathsf{g}(x),x} \wedge C(X)_{\mathsf{g}(x),\mathsf{g}(\mathsf{g}(x))}\big) \\
&\quad \vee C(X)_{s,\mathsf{g}(\mathsf{g}(x))} \\
&= \big(\bot \vee \bot \vee X_{\mathsf{gf}} \wedge \bot \wedge \top \wedge C(X)_{\mathsf{g}(x),\mathsf{g}(\mathsf{g}(x))}\big) \vee C(X)_{s,\mathsf{g}(\mathsf{g}(x))} \\
&= C(X)_{s,\mathsf{g}(\mathsf{g}(x))} \\
&= CE(X)_{s,\mathsf{g}(\mathsf{g}(x))} \vee \big(X_{\mathsf{fg}} \wedge C_1(X)_{s,\mathsf{g}(\mathsf{g}(x))}\big) \\
&= E_{y,\mathsf{g}(\mathsf{g}(x))} \vee E_{\mathsf{g}(x),\mathsf{g}(\mathsf{g}(x))} \vee E_{x,\mathsf{g}(\mathsf{g}(x))} \vee C(X)_{y,\mathsf{g}(\mathsf{g}(x))} \vee C(X)_{\mathsf{g}(x),\mathsf{g}(\mathsf{g}(x))} \\
&\quad \vee C(X)_{x,\mathsf{g}(\mathsf{g}(x))} \vee \big(X_{\mathsf{fg}} \wedge C_1(X)_{s,\mathsf{g}(\mathsf{g}(x))}\big) \\
&= \bot \vee \bot \vee \bot \vee \bot \vee C(X)_{\mathsf{g}(x),\mathsf{g}(\mathsf{g}(x))} \vee \bot \vee \big(X_{\mathsf{fg}} \wedge C_1(X)_{s,\mathsf{g}(\mathsf{g}(x))}\big) \\
&= C(X)_{\mathsf{g}(x),\mathsf{g}(\mathsf{g}(x))} \vee \big(X_{\mathsf{fg}} \wedge C(X)_{s,\mathsf{g}(x)}\big) \\
&= CE(X)_{\mathsf{g}(x),\mathsf{g}(\mathsf{g}(x))} \vee C(X)_{x,\mathsf{g}(x)} \vee \big(X_{\mathsf{fg}} \wedge C(X)_{s,\mathsf{g}(x)}\big) \\
&= E_{x,\mathsf{g}(\mathsf{g}(x))} \vee C(X)_{x,\mathsf{g}(\mathsf{g}(x))} \vee \bot \vee \big(X_{\mathsf{fg}} \wedge C(X)_{s,\mathsf{g}(x)}\big) \\
&= \bot \vee \bot \vee \big(X_{\mathsf{fg}} \wedge C(X)_{s,\mathsf{g}(x)}\big) \\
&= X_{\mathsf{fg}} \wedge \big(CE(X)_{s,\mathsf{g}(x)} \vee \big(X_{\mathsf{fg}} \wedge C_1(X)_{s,\mathsf{g}(x)}\big)\big) \\
&= X_{\mathsf{fg}} \wedge \big(E_{y,\mathsf{g}(x)} \vee E_{\mathsf{g}(x),\mathsf{g}(x)} \vee E_{x,\mathsf{g}(x)} \vee C(X)_{y,\mathsf{g}(x)} \vee C(X)_{\mathsf{g}(x),\mathsf{g}(x)} \\
&\quad \vee C(X)_{x,\mathsf{g}(x)} \vee \big(X_{\mathsf{fg}} \wedge C_1(X)_{s,\mathsf{g}(x)}\big)\big) \\
&= X_{\mathsf{fg}} \wedge \big(\bot \vee \top \vee \bot \vee \bot \vee \bot \vee \bot \vee \big(X_{\mathsf{fg}} \wedge C(X)_{s,x}\big)\big) \\
&= X_{\mathsf{fg}} \wedge \top = X_{\mathsf{fg}}
\end{aligned}$$

For the second instance we get

$$\begin{aligned}
C(X) &= C(X)_{\mathsf{f}(x),\mathsf{g}(x)} \wedge C(X)_{\mathsf{g}(x),\mathsf{f}(x)} \\
&= (CE(X)_{\mathsf{f}(x),\mathsf{g}(x)} \vee (X_{\mathsf{fg}} \wedge C(X)_{\mathsf{f}(x),x})) \\
&\quad \wedge (CE(X)_{\mathsf{g}(x),\mathsf{f}(x)} \vee (X_{\mathsf{gf}} \wedge C(X)_{\mathsf{g}(x),x})) \\
&= (E_{x,\mathsf{g}(x)} \vee C_{x,\mathsf{g}(x)} \vee (X_{\mathsf{fg}} \wedge \top)) \\
&\quad \wedge (E_{x,\mathsf{f}(x)} \vee C_{x,\mathsf{f}(x)} \vee (X_{\mathsf{gf}} \wedge \top)) = X_{\mathsf{fg}} \wedge X_{\mathsf{gf}}
\end{aligned}$$

The resulting constraints can be satisfied but they do not correspond to a strict precedence because any satisfying assignment would yield $\mathsf{f} \succ \mathsf{g}$ and $\mathsf{g} \succ \mathsf{f}$. A strict precedence is irreflexive and transitive but the relation above contradicts irreflexivity if it is closed under transitivity. The interested reader is asked to verify that $C(X,Y) = X_{\mathsf{fg}} \wedge X_{\mathsf{gf}}$ and therefore no quasi-precedence exists which shows LPO-termination of this instance. The third TRS (now computed with $Y$ variables and some details omitted) yields

$$
\begin{aligned}
&C(X,Y)_{\mathsf{div}(x,\mathsf{e}),\mathsf{i}(x)} = CE(X,Y)_{\mathsf{div}(x,\mathsf{e}),\mathsf{i}(x)} \\
&\qquad \vee (X_{\mathsf{div},\mathsf{i}} \vee (Y_{\mathsf{div},\mathsf{i}} \wedge LEX(X,Y)_{\langle x,\mathsf{e}\rangle,\langle x\rangle})) \wedge C(X,Y)_{\mathsf{div}(x,\mathsf{e}),x} \\
&\quad = \bot \vee (X_{\mathsf{div},\mathsf{i}} \vee (Y_{\mathsf{div},\mathsf{i}} \wedge \top)) \wedge \top \\
&\quad = X_{\mathsf{div},\mathsf{i}} \vee Y_{\mathsf{div},\mathsf{i}} \\
&C(X,Y)_{\mathsf{i}(\mathsf{div}(x,y)),\mathsf{div}(y,x)} = CE(X,Y)_{\mathsf{i}(\mathsf{div}(x,y)),\mathsf{div}(y,x)} \\
&\qquad \vee (X_{\mathsf{i},\mathsf{div}} \vee (Y_{\mathsf{i},\mathsf{div}} \wedge LEX(X,Y)_{\langle \mathsf{div}(x,y)\rangle,\langle y,x\rangle})) \\
&\qquad \wedge C(X,Y)_{\mathsf{i}(\mathsf{div}(x,y)),y} \wedge C(X,Y)_{\mathsf{i}(\mathsf{div}(x,y)),x} \\
&\quad = \bot \vee X_{\mathsf{i},\mathsf{div}} \vee (Y_{\mathsf{i},\mathsf{div}} \wedge \top) \wedge \top \wedge \top = X_{\mathsf{i},\mathsf{div}} \vee Y_{\mathsf{i},\mathsf{div}} \\
&C(X,Y)_{\mathsf{div}(\mathsf{div}(x,y),z),\mathsf{div}(y,\mathsf{div}(\mathsf{i}(x),z))} \\
&\quad = CE(X,Y)_{\mathsf{div}(\mathsf{div}(x,y),z),\mathsf{div}(y,\mathsf{div}(\mathsf{i}(x),z))} \vee LEX(X,Y)_{\langle \mathsf{div}(x,y),z\rangle,\langle y,\mathsf{div}(i(x),z)\rangle} \\
&\qquad \wedge C(X,Y)_{\mathsf{div}(\mathsf{div}(x,y),z),\mathsf{div}(\mathsf{i}(x),z)} \\
&\quad = \bot \vee \top \wedge C(X,Y)_{\mathsf{div}(\mathsf{div}(x,y),z),\mathsf{div}(\mathsf{i}(x),z)} \\
&\quad = CE(X,Y)_{\mathsf{div}(\mathsf{div}(x,y),z),\mathsf{div}(\mathsf{i}(x),z)} \vee LEX(X,Y)_{\langle \mathsf{div}(x,y),z\rangle,\langle \mathsf{i}(x),z\rangle} \\
&\qquad \wedge C(X,Y)_{\mathsf{div}(\mathsf{div}(x,y),z),z} \\
&\quad = \bot \vee C(X,Y)_{\mathsf{div}(x,y),\mathsf{i}(x)} \wedge \top \\
&\quad = CE(X,Y)_{\mathsf{div}(x,y),\mathsf{i}(x)} \vee (X_{\mathsf{div},\mathsf{i}} \vee (Y_{\mathsf{div},\mathsf{i}} \wedge LEX(X,Y)_{\langle x,y\rangle,\langle x\rangle})) \\
&\qquad \wedge C(X,Y)_{\mathsf{div}(x,y),x)} \\
&\quad = \bot \vee (X_{\mathsf{div},\mathsf{i}} \vee Y_{\mathsf{div},\mathsf{i}} \wedge \top) \wedge \top \\
&\quad = X_{\mathsf{div},\mathsf{i}} \vee Y_{\mathsf{div},\mathsf{i}}
\end{aligned}
$$

The conjunction of the three constraints above amounts to

$$C(X,Y) = (X_{\mathsf{div},\mathsf{i}} \vee Y_{\mathsf{div},\mathsf{i}}) \wedge (X_{\mathsf{i},\mathsf{div}} \vee Y_{\mathsf{i},\mathsf{div}})$$

and therefore the TRS admits the quasi-precedence $\mathsf{div} \sim \mathsf{i}$.

## 3.2 A Symbol Based Encoding

After computing $C(X,Y)$ one can infer a precedence from these constraints (if the necessary properties of a precedence can be satisfied). Remember that for a quasi-precedence reflexivity and transitivity have to be ensured. The idea proposed in [2] is to assign to every function symbol a positive integer value. The greater or equal than relation ($\geq$) on natural numbers then ensures that the function symbols are quasi-ordered. In fact the order is even total. Let $|\mathcal{F}| = n$. Then we are looking for a mapping $m : \mathcal{F} \to \{1,\ldots,n\}$ such that for every propositional variable $X_{fg} \in X$ we have $m(f) > m(g)$ and for $Y_{fg} \in Y$ $m(f) = m(g)$. In order to encode these constraints in propositional logic integers are represented in binary notation. To uniquely encode $n$ function symbols, $k := \lceil ld(n) \rceil$ bits are needed. Here $ld()$ denotes the logarithm to the basis 2. The next definition shows how the constraints for the $X$ (strict part) and $Y$ (equivalence part) variables can be formalised. The $k$-bit representation of $f$ is $\langle f_k, \ldots, f_1 \rangle$ with $f_k$ the most significant bit.

**Definition 12.**

$$||X_{fg}||_k = \begin{cases} (f_1 \wedge \neg g_1) & k = 1 \\ (f_k \wedge \neg g_k) \vee ((f_k \leftrightarrow g_k) \wedge ||X_{fg}||_{k-1}) & k > 1 \end{cases}$$

$$||Y_{fg}||_k = \bigwedge_{i=1}^{k} (f_i \leftrightarrow g_i)$$

After this step it is rather easy to define the whole propositional formula which is satisfiable if and only if the given TRS is (strict) LPO-terminating.

**Definition 13.**

$$B(X, Y) = C(X, Y) \wedge \bigwedge_{z \in X \cup Y} (z \leftrightarrow ||z||_k)$$

$$B(X) = C(X) \wedge \bigwedge_{z \in X} (z \leftrightarrow ||z||_k)$$

**Lemma 2.** *$B(X, Y)$ ($B(X)$) is satisfiable if and only if the TRS $\mathcal{R}$ is (strict) LPO-terminating.* □

Note that this encoding introduces new variables (i.e., after constructing $C(X, Y)$ additional variables are needed to enforce symmetry and transitivity. Let again $|\mathcal{F}| = n$. Then for every function symbol $k := \lceil ld(n) \rceil$ additional variables are added which makes a total of $\mathcal{O}(n \times ld(n))$ new variables. Nevertheless the problematic part arises from $C(X, Y)$ which typically has $\mathcal{O}(n^2)$ variables. But as this approach adds a conjunct for each propositional variable appearing in $C(X, Y)$ it thus adds $\mathcal{O}(n^2)$ conjuncts of size $\mathcal{O}(ld(n))$ (cf. Definition 12). Run time results show that these additional variables do not pose a problem for MiniSat (Section 5.1.3).

This section is concluded with a note on the symbol based encoding. Different satisfying assignments do not necessarily give rise to different total orders on the function symbols. Consider the following example:

**Example 3.** Let $C(X, Y) = X_{\mathsf{fg}} \wedge Y_{\mathsf{gh}}$. Then the three mappings $m_1, m_2, m_3 : \mathcal{F} \to \{1, 2, 3\}$ with $m_1(\mathsf{f}) = 2$, $m_1(\mathsf{g}) = m_1(\mathsf{h}) = 1$, $m_2(\mathsf{f}) = 3$, $m_2(\mathsf{g}) = m_2(\mathsf{h}) = 1$, and $m_3(\mathsf{f}) = 3$, $m_3(\mathsf{g}) = m_3(\mathsf{h}) = 2$ yield the same precedence $\mathsf{f} \succ \mathsf{g}$, $\mathsf{f} \succ \mathsf{h}$, and $\mathsf{g} \sim \mathsf{h}$.

| | | |
|---|---|---|
| $\varphi \wedge \top$ | $\to$ | $\varphi$ |
| $\varphi \wedge \bot$ | $\to$ | $\bot$ |
| $\varphi \vee \bot$ | $\to$ | $\varphi$ |
| $\varphi \vee \top$ | $\to$ | $\top$ |
| $\top \wedge \varphi$ | $\to$ | $\varphi$ |
| $\bot \wedge \varphi$ | $\to$ | $\bot$ |
| $\bot \vee \varphi$ | $\to$ | $\varphi$ |
| $\top \vee \varphi$ | $\to$ | $\top$ |

Table 1: A bunch of simplifications.

| | | |
|---|---|---|
| $\varphi \wedge (\varphi \vee \psi)$ | $\to$ | $\varphi$ |
| $\varphi \wedge (\psi \vee \varphi)$ | $\to$ | $\varphi$ |
| $(\varphi \vee \psi) \wedge \varphi$ | $\to$ | $\varphi$ |
| $(\psi \vee \varphi) \wedge \varphi$ | $\to$ | $\varphi$ |
| $\varphi \wedge \varphi$ | $\to$ | $\varphi$ |
| $\varphi \vee (\varphi \wedge \psi)$ | $\to$ | $\varphi$ |
| $\varphi \vee (\psi \wedge \varphi)$ | $\to$ | $\varphi$ |
| $(\varphi \wedge \psi) \vee \varphi$ | $\to$ | $\varphi$ |
| $(\psi \wedge \varphi) \vee \varphi$ | $\to$ | $\varphi$ |
| $\varphi \vee \varphi$ | $\to$ | $\varphi$ |

Table 2: Some more simplifications.

# 4 Optimisations

## 4.1 Modifications

In addition to the constraint encoding in [2] some slight modifications have been implemented. The propositional formula resulting from the LPO encoding typically contains many occurrences of $\bot$ and $\top$. What happens when the simplifications of Table 1 are performed before testing for satisfiability? Is it even better if some more simplifications like the ones in Table 2 are added? Although not explicitly mentioned, the implementation of [2] incorporates the equivalences of Table 1. The simplifications of Tables 1 and 2 have been integrated in our implementation. Run time results show that the first bunch of simplifications should really be employed whereas the second one does not give an enormous speedup. The reason why the simplifications of Table 1 are so effective is that they usually reduce the number of propositional variables. For example $X_{\mathsf{fg}} \wedge X_{\mathsf{gh}} \wedge X_{\mathsf{hk}} \wedge \bot$ is reduced to $\bot$. The author of this report is convinced that some further simplifications would be useful in order to reduce the number of subformulas, which is relevant for the transformation to CNF (cf. Definition 14).

| | | | | | | |
|---|---|---|---|---|---|---|
| (1) | $\varphi \wedge \top$ | $\to$ | $\varphi$ | $\varphi \vee \top$ | $\to$ | $\top$ |
| | $\top \wedge \varphi$ | $\to$ | $\varphi$ | $\top \vee \varphi$ | $\to$ | $\top$ |
| | $\varphi \wedge \bot$ | $\to$ | $\bot$ | $\varphi \vee \bot$ | $\to$ | $\varphi$ |
| | $\bot \wedge \varphi$ | $\to$ | $\bot$ | $\bot \vee \varphi$ | $\to$ | $\varphi$ |
| (2) | $\varphi \to \psi$ | $\to$ | $\neg\varphi \vee \psi$ | $\varphi \leftrightarrow \psi$ | $\to$ | $\varphi \wedge \psi \vee \neg\varphi \wedge \neg\psi$ |
| (3) | $\neg(\varphi \wedge \psi)$ | $\to$ | $\neg\varphi \vee \neg\psi$ | $\neg(\varphi \vee \psi)$ | $\to$ | $\neg\varphi \wedge \neg\psi$ |
| (4) | $\neg\neg\varphi$ | $\to$ | $\varphi$ | | | |
| (5) | $\varphi \vee (\psi \wedge \chi)$ | $\to$ | $(\varphi \vee \psi) \wedge (\varphi \vee \chi)$ | | | |
| | $(\varphi \wedge \psi) \vee \chi$ | $\to$ | $(\varphi \vee \chi) \wedge (\psi \vee \chi)$ | | | |

Table 3: The standard transformation into CNF.

## 4.2 Constructing CNFs Efficiently

SAT solvers typically expect their input in conjunctive normal form but for the majority of the TRSs the formula $B(X,Y)$ is too large for the standard translation which consists of the five steps depicted in Table 3. The problem there is that the resulting CNF may be exponentially larger than the input formula because the two rules of step (5) duplicate one of the variables. In [11] Tseitin proposed a transformation which is linear in the size of the input formula. The price for linearity is paid with introducing new variables. As a consequence, Tseitin's transformation does not produce an equivalent formula. E.g., a tautology may no longer be a tautology because of the new variables. But it preserves and reflects satisfiability. The basic idea of this transformation is simple: In order to transform the formula $\varphi$ introduce for every non-atomic subformula $\psi$ a new variable $p_\psi$. Atoms $\psi$ are identified with $p_\psi$. The translation of the formula is presented in the definition below. $NASub(\varphi)$ denotes all non-atomic subformulas of $\varphi$ and $*$ represents all binary connectives.

**Definition 14.**

$$Tseitin(\varphi) = p_\varphi \wedge \bigwedge_{\substack{\psi \in NASub(\varphi) \\ \psi = \psi_1 * \psi_2}} (p_\psi \leftrightarrow (p_{\psi_1} * p_{\psi_2})) \wedge \bigwedge_{\substack{\psi \in NASub(\varphi) \\ \psi = \neg\psi_1}} (p_\psi \leftrightarrow \neg p_{\psi_1})$$

The attentive reader may be puzzled because the definition above does not produce a CNF. However, every of the conjuncts above can be represented in CNF using at most four clauses (cf. Table 4). This section is concluded with a simple example for Tseitin's transformation.

| | | |
|---|---|---|
| $\chi \leftrightarrow (\varphi \wedge \psi)$ | $\to$ | $(\neg\varphi \vee \neg\psi \vee \chi) \wedge (\varphi \vee \neg\chi) \wedge (\psi \vee \neg\chi)$ |
| $\chi \leftrightarrow (\varphi \vee \psi)$ | $\to$ | $(\varphi \vee \psi \vee \neg\chi) \wedge (\neg\varphi \vee \chi) \wedge (\neg\psi \vee \chi)$ |
| $\chi \leftrightarrow (\varphi \to \psi)$ | $\to$ | $(\neg\varphi \vee \psi \vee \neg\chi) \wedge (\varphi \vee \chi) \wedge (\neg\psi \vee \chi)$ |
| $\chi \leftrightarrow (\varphi \leftrightarrow \psi)$ | $\to$ | $(\neg\varphi \vee \neg\psi \vee \neg\chi) \wedge (\varphi \vee \psi \vee \chi)$ $\wedge (\varphi \vee \neg\psi \vee \neg\chi) \wedge (\neg\varphi \vee \psi \vee \neg\chi)$ |
| $\chi \leftrightarrow (\neg\varphi)$ | $\to$ | $(\varphi \vee \chi) \wedge (\neg\varphi \vee \neg\chi)$ |

Table 4: Some equivalences.

**Example 4.** Let $\varphi = (q \wedge \neg r) \vee s$. Then

$$NASub(\varphi) = \{\underbrace{\varphi}_{p_\varphi}, \underbrace{q \wedge \neg r}_{p_{\psi_1}}, \underbrace{\neg r}_{p_{\psi_2}}\}$$

and

$$Tseitin(\varphi) = p_\varphi \wedge (p_\varphi \leftrightarrow (p_{\psi_1} \vee s)) \wedge (p_{\psi_1} \leftrightarrow (q \wedge p_{\psi_2})) \wedge (p_{\psi_2} \leftrightarrow \neg r).$$

### 4.2.1 Some Remarks

In order to get a shorter representation one can try to replace $\leftrightarrow$ in Definition 14 by $\rightarrow$. This is a bit dangerous as the following example demonstrates.

**Example 5.** Let $\varphi$ be the propositional formula $\neg(p \vee \neg p)$ and therefore

$$NASub(\varphi) = \{\underbrace{\varphi}_{p_\varphi}, \underbrace{(p \vee \neg p)}_{p_{\psi_1}}, \underbrace{\neg p}_{p_{\psi_2}}\}.$$

The transformed formula is satisfiable although the original formula $\neg(p \vee \neg p)$ is unsatisfiable. Below an example satisfying assignment is provided.

$$\begin{aligned} Tseitin'(\varphi) = {} & \underbrace{p_\varphi}_{\mathsf{T}} \wedge (\underbrace{p_\varphi}_{\mathsf{T}} \rightarrow \neg \underbrace{p_{\psi_1}}_{\mathsf{F}}) \wedge (\underbrace{p_{\psi_1}}_{\mathsf{F}} \rightarrow \underbrace{p}_{\mathsf{F}} \vee \underbrace{p_{\psi_2}}_{\mathsf{F}}) \\ & \wedge (\underbrace{p_{\psi_2}}_{\mathsf{F}} \rightarrow \neg \underbrace{p}_{\mathsf{F}}) \end{aligned}$$

In the sequel it will become clear that this problem is caused by negations not applied to atoms. Therefore we restrict ourselves to negation normal form (NNF), i.e., negations are only allowed in front of atoms. Steps (1) – (4) of Table 3 are unproblematic as far as complexity is concerned and transform the input into NNF. Afterwards the changed transformation comes into action. It introduces a new formula for each non-literal subformula instead of every non-atomic subformula. *NLSub*($\varphi$) denotes all non-literal subformulas of $\varphi$.

**Definition 15.**

$$Tseitin'(\varphi) = p_\varphi \wedge \bigwedge_{\substack{\psi \in NLSub(\varphi) \\ \psi = \psi_1 * \psi_2}} (p_\psi \rightarrow (p_{\psi_1} * p_{\psi_2}))$$

Valid precedences correspond to satisfying assignments of the formula. Therefore it is important not to lose a precedence and, moreover, to get valid ones only. The former is to say that every satisfying assignment of $\varphi$ can be extended (note that there are additional variables) to a satisfying assignment of $Tseitin'(\varphi)$. The latter expresses that every satisfying assignment for the transformed formula should also satisfy the original one.

**Lemma 3.** *Let $\alpha$ be an assignment with $\alpha(\varphi) = \mathsf{T}$. Then $\alpha$ can be extended to some $\alpha'$ such that $\alpha'(Tseitin'(\varphi)) = \mathsf{T}$.*

*Proof.* As the result holds for the original transformation the same assignment also satisfies $Tseitin'(\varphi)$ because its formulation is weaker. □

**Lemma 4.** *Let $\alpha$ be an assignment. If $\alpha(Tseitin'(\varphi)) = \mathsf{T}$ then $\alpha(\varphi) = \mathsf{T}$.*

*Proof.* The proof is done by induction on the number of non-literal subformulas. As an abbreviation define:

$$conj(\varphi) = \begin{cases} \varphi & \text{if } \varphi \text{ is a literal} \\ \bigwedge\limits_{\substack{\psi \in NLSub(\varphi) \\ \psi = \psi_1 * \psi_2}} (p_\psi \to (p_{\psi_1} * p_{\psi_2})) & \text{otherwise} \end{cases}$$

The base case amounts to verifying that for all literals the statement holds.

- If $\varphi$ is a literal, then $Tseitin'(\varphi) = \varphi$ by definition and the result follows immediately.

In the inductive step we must verify that for all binary operators (as NNF is considered there are only $\wedge$ and $\vee$) the result holds.

- $\varphi = \psi_1 \wedge \psi_2$ :

$$\begin{aligned} & \alpha(Tseitin'(\varphi)) = \mathsf{T} \\ \Longleftrightarrow\ & \alpha(p_\varphi \wedge conj(\varphi)) = \mathsf{T} \\ \Longleftrightarrow\ & \alpha(p_\varphi \wedge (p_\varphi \to p_{\psi_1} \wedge p_{\psi_2}) \wedge conj(\psi_1) \wedge conj(\psi_2)) = \mathsf{T} \\ \Longleftrightarrow\ & \alpha(p_\varphi \wedge (p_\varphi \to p_{\psi_1} \wedge p_{\psi_2})) = \mathsf{T} \text{ and } \alpha(conj(\psi_1) \wedge conj(\psi_2)) = \mathsf{T} \\ \Longrightarrow\ & \alpha(p_{\psi_1}) = \mathsf{T} \text{ and } \alpha(p_{\psi_2}) = \mathsf{T} \text{ and } \alpha(conj(\psi_1)) = \mathsf{T} \\ & \text{and } \alpha(conj(\psi_2)) = \mathsf{T} \\ \Longleftrightarrow\ & \alpha(p_{\psi_1} \wedge conj(\psi_1)) = \mathsf{T} \text{ and } \alpha(p_{\psi_2} \wedge conj(\psi_2)) = \mathsf{T} \\ \Longleftrightarrow\ & \alpha(Tseitin'(\psi_1)) = \mathsf{T} \text{ and } \alpha(Tseitin'(\psi_2)) = \mathsf{T} \\ \overset{\text{IH}}{\Longleftrightarrow}\ & \alpha(\psi_1) = \mathsf{T} \text{ and } \alpha(\psi_2) = \mathsf{T} \\ \Longleftrightarrow\ & \alpha(\psi_1 \wedge \psi_2) = \mathsf{T} \\ \Longleftrightarrow\ & \alpha(\varphi) = \mathsf{T} \end{aligned}$$

- $\varphi = \psi_1 \vee \psi_2$:
  similar

□

Unfortunately the encoding of [2] does not produce a propositional formula in NNF (the one in [6] does!). Test results show that it is cheaper to use Tseitin's transformation in its original definition because the transformation to negation normal form – although linear – is too expensive. But one refinement to Tseitin's transformation can be made (and is implemented) which really speeds up the whole process: Consider only non-literal subformulas instead of non-atomic ones. The transformed formula is then not necessarily in CNF because atoms may occur with more than one negation. Removal of those negations is computationally cheaper than considering also negated atoms as subformulas. Also note that the propositional formula $B(X, Y)$ already consists of some conjunctions, say $c_1 \wedge \cdots \wedge c_n$. Then $Tseitin(c_1) \wedge \cdots \wedge Tseitin(c_n)$ is computed a bit faster than $Tseitin(c_1 \wedge \cdots \wedge c_n)$. Remember that the $n$ is at least the number of rewrite rules plus function symbols occurring in $B(X, Y)$.

In [12] a different (from the one in Definition 15) formulation is presented where $\leftrightarrow$ in Definition 14 is also replaced by $\rightarrow$. In this version the transformed formula might have more satisfying assignments if the input formula is satisfiable. It would be sufficient for LPO-termination if the transformed formula is satisfiable (because also in this transformation the output formula is satisfiable if and only if the input formula was) but a satisfying assignment can no longer be used for reading off a valid precedence because not every satisfying assignment of the output also satisfies the input.

# 5 Experimental Results

In this section the standard LPO implementation of T┬T (referred to as `TTT`) is compared with two of the three different atom based encodings described in [6] using the implementation of [13] (referred to `BDD2` and `BDD3`) and a new implementation (referred to as `SAT`) of the symbol based encoding due to [2] and described above. Of course the comparison is a bit unfair since the atom based encodings use BDD techniques to test satisfiability whereas the symbol based approach interfaces the high-end SAT solver MiniSat. For the test benches a database of 773 TRSs [5] is considered and the results of interesting examples are described in more detail. All tests were performed on `cl2-informatik.uibk.ac.at`, a server with two Intel® Xeon™ processors running at a CPU rate of 2.40GHz. The system memory is 512MB in total.
The abbreviations `TO` (for timeout, 10 seconds) and `SO` (for stack overflow) are used. The simplifications of Tables 1 and 2 together with the most effective variable order `wao` (cf. [13]) were employed. Times in the tables are in seconds and include reading the input file, deciding LPO-termination, and printing a precedence if there exists one.

| TRS | `TTT` | `BDD2` | `BDD3` | `SAT` |
|---|---|---|---|---|
| Cime_tree | `TO` | 0.044 | 0.043 | 0.025 |
| currying_AG01_3.10 | `TO` | 0.056 | 0.054 | 0.018 |
| currying_AG01_3.13 | `TO` | 0.083 | 0.082 | 0.018 |
| currying_Ste92_hydra | `TO` | 0.089 | 0.052 | 0.015 |
| HM_t005 | `TO` | 0.447 | 0.449 | 0.449 |
| SK90_4.47 | `TO` | 0.032 | 0.032 | 0.020 |
| various_14 | `TO` | 0.061 | 0.061 | 0.042 |
| Zantema_z30 | `TO` | 0.091 | 0.091 | 0.091 |

Table 5: Problematic instances for `TTT`.

| TRS | BDD3 | BDD2 | TTT | SAT |
|---|---|---|---|---|
| AProVE_AAECC-ring | 0.10 | TO | 0.06 | 0.12 |
| Cime_quick | TO | 0.07 | 0.01 | 0.02 |
| HM_t009 | TO | TO | 0.07 | 0.13 |
| Rubio_enno | *0.21* | 0.06 | 0.00 | 0.02 |
| Rubio_wst99 | *0.41* | 0.93 | 0.01 | 0.02 |
| secret2005_cime3 | *0.13* | 6.94 | 0.02 | 0.06 |
| TRCSR_Ex1_2_AEL03_C | *0.33* | TO | 0.12 | 0.16 |
| TRCSR_Ex1_2_AEL03_GM | *0.11* | TO | 0.03 | 0.07 |
| TRCSR_Ex14_AEGL02_C | TO | TO | 0.03 | 0.03 |
| TRCSR_Ex14_AEGL02_FR | *0.21* | 0.22 | 0.01 | 0.02 |
| TRCSR_Ex1_GL02a_C | TO | TO | 0.03 | 0.04 |
| TRCSR_Ex1_GM03_C | *0.69* | TO | 0.03 | 0.04 |
| TRCSR_Ex1_Luc02b_C | *3.29* | 1.87 | 0.02 | 0.03 |
| TRCSR_Ex26_Luc03b_C | TO | TO | 0.06 | 0.06 |
| TRCSR_Ex26_Luc03b_FR | 0.90 | TO | 0.04 | 0.05 |
| TRCSR_Ex2_Luc02a_C | *6.29* | TO | 0.06 | 0.14 |
| TRCSR_Ex2_Luc03b_C | *0.79* | TO | 0.03 | 0.08 |
| TRCSR_Ex3_2_Luc97_C | TO | TO | 0.03 | 0.04 |
| TRCSR_Ex3_2_Luc97_FR | 0.44 | TO | 0.03 | 0.04 |
| TRCSR_Ex3_3_25_Bor03_C | *0.71* | TO | 0.04 | 0.04 |
| TRCSR_Ex3_3_25_Bor03_FR | 0.11 | TO | 0.03 | 0.03 |
| TRCSR_Ex4_7_37_Bor03_C | *3.25* | TO | 0.06 | 0.06 |
| TRCSR_Ex49_GM04_C | *0.93* | TO | 0.05 | 0.04 |
| TRCSR_Ex5_7_Luc97_C | *0.12* | TO | 0.09 | 0.12 |
| TRCSR_Ex5_7_Luc97_FR | *0.12* | TO | 0.04 | 0.06 |
| TRCSR_Ex5_7_Luc97_GM | *0.50* | TO | 0.03 | 0.07 |
| TRCSR_Ex5_7_Luc97_Z | *0.13* | TO | 0.04 | 0.05 |
| TRCSR_Ex6_15_AEL02_C | *8.13* | TO | 0.18 | 0.25 |
| TRCSR_Ex6_15_AEL02_FR | *1.00* | TO | 0.11 | 0.09 |
| TRCSR_Ex6_15_AEL02_GM | TO | TO | 0.04 | 0.12 |
| TRCSR_Ex6_15_AEL02_Z | *1.03* | TO | 0.09 | 0.08 |
| TRCSR_Ex7_BLR02_C | TO | TO | 0.04 | 0.05 |
| TRCSR_Ex8_BLR02_C | TO | TO | 0.04 | 0.04 |
| TRCSR_Ex9_BLR02_C | *0.34* | 5.82 | 0.05 | 0.04 |
| TRCSR_ExAppendixB_AEL03_C | *0.57* | TO | 0.16 | 0.18 |
| TRCSR_ExIntrod_GM01_C | *3.14* | TO | 0.04 | 0.04 |
| TRCSR_ExIntrod_GM04_C | *6.31* | TO | 0.02 | 0.04 |
| TRCSR_ExIntrod_GM99_C | *5.68* | TO | 0.10 | 0.08 |
| TRCSR_ExIntrod_GM99_FR | *0.30* | TO | 0.03 | 0.03 |
| TRCSR_ExIntrod_GM99_GM | *0.15* | TO | 0.03 | 0.06 |
| TRCSR_ExIntrod_Zan97_C | TO | TO | 0.06 | 0.06 |
| TRCSR_ExSec11_1_Luc02a_C | TO | TO | 0.08 | 0.08 |

Table 6: Instances where BDD approach fails.

| TRS | SAT | BDD2 | BDD3 | TTT |
|---|---|---|---|---|
| AProVE_AAECC-ring | 0.117 | TO | 0.081 | 0.062 |
| Cime_mucrl1 | 0.296 | 0.298 | 0.296 | 0.932 |
| HM_t005 | 0.449 | 0.447 | 0.449 | TO |
| HM_t009 | 0.123 | TO | TO | 0.062 |
| TRCSR_Ex1_2_AEL03_C | 0.150 | TO | *0.332* | 0.121 |
| TRCSR_Ex5_7_Luc97_C | 0.117 | TO | TO | 0.089 |
| TRCSR_Ex6_15_AEL02_C | 0.233 | TO | *8.134* | 0.171 |
| TRCSR_Ex6_15_AEL02_GM | 0.109 | TO | TO | 0.038 |
| TRCSR_ExAppendixB_AEL03_C | 0.172 | TO | *0.572* | 0.151 |
| Zantema_z30 | 0.091 | 0.091 | 0.091 | TO |

Table 7: The 10 hardest problems for the symbol based encoding.

### 5.1 Comparing the Three Approaches

In this section the most expensive instances for each approach are discussed. Currently there is no implementation for quasi-LPO-termination following the atomic encoding idea. Therefore only results concerning strict LPO-termination are reported here. A concise comparison for quasi-LPO-termination between the symbol based encoding and TTT can be found in [2].

#### 5.1.1 Problematic TTT Instances

Here the eight instances where TTT could not decide LPO-termination within a timeout of ten seconds are considered. In Table 5 the execution times of the alternative approaches are shown. The BDD approach is much better than TTT (for these instances) and SAT is even faster.

#### 5.1.2 Problematic Instances for the Atom Based Encoding

In Table 6 the instances which could not be handled by BDD2 or BDD3 (either because of timeout or stack overflow) are compared with the results of the TTT and SAT implementations. There are three different types of table entries. Times for a successful decision of LPO-termination are written in roman font. *Italic* numbers indicate the time after which a stack overflow occurred and TO means that LPO-termination could not be decided within the given timeout. Note that stack overflows only occurred using BDD3. Whereas both BDD approaches usually have difficulties with the same instances these ones seem to cause no problem for TTT and SAT.

#### 5.1.3 Problematic Instances for the Symbol Based Encoding

Surprisingly there are no really problematic TRSs for SAT. No timeouts or stack overflows occurred and the run time results are unequivocal. Table 7 shows the ten hardest instances for the symbol based approach. Just note that although the SAT approach seems favourable (cf. Section 5.2) BDD3 is equally fast for the two hardest SAT problems. The reason is that for these two systems almost all execution time is used for generating the LPO constraints.

| method | cpu time | timeouts | stack overflows |
|---|---|---|---|
| BDD3 | 116 | 4 | 31 |
| BDD2 | 466 | 36 | 0 |
| TTT | 117 | 8 | 0 |
| SAT | 8 | 0 | 0 |

Table 8: The three approaches tested on 773 TRSs.

### 5.2 A Global View

Until now only separated instances have been considered. This is the right place to apply the different methods to all of the 773 TRSs. Table 8 compares the results of the BDD approach with TTT and SAT. The format of the table entries is cpu time/timeouts/stack overflows. So the first entry should be low (execution time) and also the sum of the second and third (where decision of LPO-termination fails). Comparing the results of both BDD approaches the big difference is the 31 stack overflows caused by BDD3. As explained in [13] these typically occur within the first second when computing the cycles whereas BDD2 computes minimal paths and minimal cycles until the timeout is reached. (So $31 \times 10 = 310$ which is roughly the difference between the two execution times). The last two lines list the results for TTT and SAT. The BDD approach does not perform so well here because a subclass of TRSs in the database causes problems. This sub class is formed by some of the TRSs which fit the naming TRCSR_*. The reason for this is the number of cycles in the domain graph. Without these instances the BDD approach performs much better (see [13]). SAT can handle all instances (no timeouts, no stack overflows) and thus is preferable.

Finally let us compare our SAT implementation with the results of the original poSAT implementation of [2]. Table 9 presents the 773 instances tested with SAT and compares the results with the 751 instances tested with poSAT. Remember that the numbers were produced on different machines and thus it might be dangerous to compare them directly.

| | poSAT | SAT |
|---|---|---|
| Total | 9.112 | 7.666 |
| Average | 0.012 | 0.010 |
| Max | 0.450 | 0.449 |

(a) strict LPO

| | poSAT | SAT |
|---|---|---|
| Total | 10.428 | 10.690 |
| Average | 0.014 | 0.014 |
| Max | 1.169 | 0.532 |

(b) quasi LPO

Table 9: 751 instances for poSAT versus 773 instances for SAT.

| TRS | poSAT | SAT |
|---|---|---|
| HM_t005 | 0.450 | 0.443 |
| Cime_mucrl1 | 0.294 | 0.313 |
| currying_AG01_No_3.13 | 0.127 | 0.019 |

Table 10: Run times for costly to encode instances.

### 5.3 General Remarks

While performing the tests for the atom based encoding two major problems arose. The one with the stack overflow and the timeouts. The former is caused by the cycle computation needed for BDD3. The current algorithm to obtain all cycles can surely be improved but will still remain computationally expensive. The troubles with the timeout splits into two subproblems. Either the constraint formula $B(X, Y)$ cannot be computed within the allowed time or the timeout occurs while building the BDD. In the first case no refinement will help but in the second extra memory may help. The reason is that if the BDD becomes bigger the whole amount of 512MB memory is allocated within some seconds and because of intensive swapping most computations seem to last forever.

Concerning the symbol based approach there won't be much refinements for further speedup. The authors of [2] do memoization when computing the LPO constraint $C(X, Y)$ because for some instances the same test $s \succ_{\mathrm{lpo}} t$ is performed over and over again. Maybe that is just because the definition of $\succ_{\mathrm{lpo}}$ is a bit less efficient compared to the one presented here. Our implementation does no memoization and produces more or less equal results. Table 10 lists three examples where $C(X, Y)$ simplifies to $\top$. That is to say that almost the whole execution time is spend on computing the LPO encoding of the instance.

## 6 Comments on the Paper

Reference [6] formed the prerequisites of this work. Afterwards the focus shifted to the alternative encoding presented in [2]. With the first paper as preparation the second one is easy to understand. The examples given help to grasp the definitions. Also the algorithm is tested on a large database of TRSs which allows to draw valid conclusions about the results which are then presented in detail. So the impression of the paper is rather good but some details should have been explained more precisely (e.g., Tseitin's transformation, interface with MiniSat). Therefore re-implementing the algorithm was more work than anticipated and at first the times obtained in [2] could not be reproduced. Furthermore it is unclear which transformation to CNF actually was applied. Either the one in [11], which has the advantage that precedences can be inferred and is also the one applied here, or the one in [12] which might be faster but does not allow one to conclude a valid precedence.

# 7 Future Work

As addressed in Appendix B the current interface for MiniSat is kind of minimalistic. To accelerate the whole process more C++ data types and maybe also classes should be interfaced. Also the encoding itself could be optimised a bit. Equivalences of the form $X_{fg} \leftrightarrow \neg X_{gf}$ as well as $Y_{fg} \leftrightarrow Y_{gf}$ might reduce the number of propositional variables in $C(X, Y)$ by a factor 2 and therefore fewer constraints have to be added for ordering the function symbols. The alternative transformation to CNF in [12] might also give some speedup. An extension of the current work to MPO (Multiset Path Order) might be doable without a big effort – just design an encoding for the MPO constraints. Although MPO can solve instances which LPO cannot it is somehow weaker than LPO which can decide termination of more TRSs in the database [5] (strict/quasi-LPO can prove 128/132 TRSs terminating, strict MPO only 93, 88 of these instances can be proved by both). Relating this approach to the dependency pair method ([1]) where finding an appropriate argument filtering is one of the main bottlenecks ([4]) may be worth a consideration. How to efficiently encode the constraints for that problem needs some further investigation.

# 8 Summary

In this seminar report the main ideas of a symbol based encoding for LPO-termination proposed in [2] are explained. Furthermore an implementation in `OCaml` has been produced and its run time results are compared to the poSAT implementation of [2], the standard LPO implementation of T<sub>T</sub>T and a BDD implementation which follows a different approach ([6]). The run time results are unequivocal, i.e., the symbol based encoding is the clear winner.

# Acknowledgements

Thanks to Aart Middeldorp for supplying many ideas as well as for the excellent guidance through the whole seminar. Nao Hirokawa and Andrzej Janikowski donated the implementation of T<sub>T</sub>T and an OBDD module respectively. Christian Sternagel gave some assistance for interfacing C++ with `OCaml`. Anna Maria Scheiring helped accessing the original paper of Tseitin [11].

# A LPO is a Simplification Order

In the sequel we show that $\succ_{\text{lpo}}$ as defined in Section 2.2 indeed is a simplification order and therefore a sufficient condition for termination. First we have to prove some properties of the relation $\sim_{\text{lpo}}$.

**Lemma 5.** $\sim_{\text{lpo}}$ *is an equivalence relation that is closed under substitutions.*

*Proof.* As reflexivity, symmetry, and transitivity are obvious we only show closure under substitutions, i.e., $s \sim_{\text{lpo}} t$ implies $s\sigma \sim_{\text{lpo}} t\sigma$ for all terms $s, t$, and substitutions $\sigma$. Assume that $s \sim_{\text{lpo}} t$. If $s = t$ then clearly $s\sigma = t\sigma$ and therefore $s\sigma \sim_{\text{lpo}} t\sigma$. For the other case in the definition we do induction on $||s|| + ||t||$. In the base case we have $||s|| = ||t|| = 0$. Then $s \sim_{\text{lpo}} t$ amounts to $s = t$ and therefore $s\sigma = t\sigma$ which implies $s\sigma \sim_{\text{lpo}} t\sigma$. In the inductive step assume that $s' \sim_{\text{lpo}} t'$ implies $s'\sigma \sim_{\text{lpo}} t'\sigma$ for all $s', t'$ with $||s'|| + ||t'|| < k$ ($k > 0$) and let $||s|| + ||t|| = k$. Assuming $s \sim_{\text{lpo}} t$ yields $s = f(s_1, \ldots, s_m)$ and $t = g(t_1, \ldots, t_m)$ with $f \sim g$ and $s_i \sim_{\text{lpo}} t_i$ for all $1 \leq i \leq m$. The induction hypothesis yields $s_i\sigma \sim_{\text{lpo}} t_i\sigma$ for all $1 \leq i \leq m$ and thus $s\sigma \sim_{\text{lpo}} t\sigma$. □

**Lemma 6.** *The inclusions* $\sim_{\text{lpo}} \cdot \succ_{\text{lpo}} \subseteq \succ_{\text{lpo}}$ *and* $\succ_{\text{lpo}} \cdot \sim_{\text{lpo}} \subset \succ_{\text{lpo}}$ *hold.*

*Proof.* We only prove $\sim_{\text{lpo}} \cdot \succ_{\text{lpo}} \subseteq \succ_{\text{lpo}}$, i.e., if $s \sim_{\text{lpo}} t$ and $t \succ_{\text{lpo}} u$ then $s \succ_{\text{lpo}} u$, because the other inclusion is similar. The proof is done by induction on $||s|| + ||t|| + ||u||$. Let $s = f(s_1, \ldots, s_m)$ and $t = g(t_1, \ldots, t_m)$ with $f \sim g$. In the base case $||s|| = ||t|| = 1$ and $||u|| = 0$. $t \succ_{\text{lpo}} u$ can only be by $\langle 3, i\rangle$ and as $u = x$ for some $x \in \mathcal{V}$ we have $t_i = u$. Furthermore $s_i = t_i = u$ yields $s \succ_{\text{lpo}} u$ by $\langle 3, i\rangle$. For the inductive step assume that $s' \sim_{\text{lpo}} t' \succ_{\text{lpo}} u'$ implies $s' \succ_{\text{lpo}} u'$ for all terms $s', t', u'$ with $||s'||+||t'||+||u'|| < k$ ($k > 2$) and $||s||+||t||+||u|| = k$. Consider the three cases

(1) Let $t \succ_{\text{lpo}} u$ by case (1). Then $u = h(u_1, \ldots, u_p)$ with $g \sim h$. First consider the case where $t_j \sim_{\text{lpo}} u_j$ for all $1 \leq j < i$, $t_i \succ_{\text{lpo}} u_i$, and $t \succ_{\text{lpo}} u_j$ for all $i < j \leq p$. $s \sim_{\text{lpo}} t$ implies $s_j \sim_{\text{lpo}} t_j$ for all $1 \leq j \leq m$ and by transitivity of $\sim_{\text{lpo}}$ also $s_j \sim_{\text{lpo}} u_j$ for all $1 \leq j < i$. $s_i \sim_{\text{lpo}} t_i \succ_{\text{lpo}} u_i$ implies $s_i \succ_{\text{lpo}} u_i$ by the induction hypothesis and $s \sim_{\text{lpo}} t \succ_{\text{lpo}} u_j$ implies $s \succ_{\text{lpo}} u_j$ for all $i < j \leq p$ by the induction hypothesis. Therefore $s \succ_{\text{lpo}} u$ $\langle 1, i\rangle$. In the other case we have $n > p$ and $t_i \sim_{\text{lpo}} u_i$ for all $1 \leq i \leq p$. Since $s \sim_{\text{lpo}} t$ $m = n$ and $s_i \sim_{\text{lpo}} t_i$ for all $1 \leq i \leq n$. Transitivity of $\sim_{\text{lpo}}$ yields $s_i \sim_{\text{lpo}} u_i$ for all $1 \leq i \leq p$ which together with $m > p$ proves $s \succ_{\text{lpo}} t$ $\langle 1\rangle$.

(2) $t \succ_{\text{lpo}} u$ $\langle 2\rangle$. Then $u = h(u_1, \ldots, u_p)$ with $g \succ h$ and $t \succ_{\text{lpo}} u_j$ for all $1 \leq j \leq p$. $s \succ_{\text{lpo}} u_j$ can be obtained by applying the induction hypothesis to $s \sim_{\text{lpo}} t \succ_{\text{lpo}} u_j$ for $1 \leq j \leq p$ and therefore $s \succ_{\text{lpo}} u$ $\langle 2\rangle$ because $f \succ h$.

(3) $t \succ_{\text{lpo}} u$ $\langle 3, i\rangle$. Then either $t_i \sim_{\text{lpo}} u$ or $t_i \succ_{\text{lpo}} u$. For the former we get $s_i \sim_{\text{lpo}} t_i \sim_{\text{lpo}} u$ and by transitivity of $\sim_{\text{lpo}}$ also $s_i \sim_{\text{lpo}} u$ and thus $s \succ_{\text{lpo}} u$ $\langle 3, i\rangle$. The latter yields $s_i \sim_{\text{lpo}} t_i \succ_{\text{lpo}} u$ and thus $s_i \succ_{\text{lpo}} u$ by the induction hypothesis and finally $s \succ_{\text{lpo}} u$ $\langle 3, i\rangle$.

$\square$

Next we show that $\succ_{\text{lpo}}$ is a simplification order. The proofs are slightly adapted from [8] where only strict precedences are considered.

**Lemma 7.** *$\succ_{\text{lpo}}$ is a rewrite relation, i.e., closed under contexts and substitutions.*

*Proof.* Closure under contexts: It suffices to show that $s \succ_{\text{lpo}} t$ implies $C[s] \succ_{\text{lpo}} C[t]$ for all contexts of the form $f(u_1, \dots, \square, \dots, u_m)$. Let $\square$ be the $i$-th argument of $C$. Assume $s \succ_{\text{lpo}} t$. We have to show $C[s] \succ_{\text{lpo}} C[t]$. $C[s] = f(u_1, \dots, s, \dots, u_m) \succ_{\text{lpo}} f(u_1, \dots, t, \dots, u_m)$ $\langle 1, i \rangle$ because $u_j \sim_{\text{lpo}} u_j$ (even $u_j = u_j$) for all $1 \leq j < i$, $s \succ_{\text{lpo}} t$ by assumption, and $C[s] \succ_{\text{lpo}} t_j$ for all $i < j \leq m$.

Closure under substitutions: By induction on $||s|| + ||t||$ we show that $s \succ_{\text{lpo}} t$ implies $s\sigma \succ_{\text{lpo}} t\sigma$. Assume $s \succ_{\text{lpo}} t$. Therefore let $s = f(s_1, \dots, s_m)$. In the base case $||s|| = 1$ and $||t|| = 0$. As $t \in \mathcal{V}$ there must be an $i$ ($1 \leq i \leq m$) with $s_i \sim_{\text{lpo}} t$ (even $s_i = t$). Then clearly $s_i\sigma \sim t\sigma$ and therefore $s\sigma \succ_{\text{lpo}} t\sigma$ by $\langle 3, i \rangle$. For the inductive step assume that $s' \succ_{\text{lpo}} t'$ implies $s'\sigma \succ_{\text{lpo}} t'\sigma$ for all substitutions $\sigma$ and terms $s', t'$ with $||s'|| + ||t'|| < k$ ($k > 1$) and let $||s|| + ||t|| = k$. Furthermore $s = f(s_1, \dots, s_m)$ and $t = g(t_1, \dots, t_n)$. We distinguish three cases

- If $s \succ_{\text{lpo}} t$ $\langle 1, i \rangle$ or $\langle 1 \rangle$ then $f \sim g$. First consider the case where $s_j \sim_{\text{lpo}} t_j$ for all $1 \leq j < i$, $s_i \succ_{\text{lpo}} t_i$, and $s \succ_{\text{lpo}} t_j$ for all $i < j \leq n$. Lemma 5 yields $s_j\sigma \sim_{\text{lpo}} t_j\sigma$ for all $1 \leq j < i$ and the induction hypothesis yields $s_i\sigma \succ_{\text{lpo}} t_i\sigma$, and $s\sigma \succ_{\text{lpo}} t_j\sigma$ for all $i < j \leq n$. Consequently, $s\sigma \succ_{\text{lpo}} t\sigma$ $\langle 1, i \rangle$. In the other case where $m > n$ and $s_i \sim_{\text{lpo}} t_i$ for all $1 \leq i \leq n$ Lemma 5 yields $s_i\sigma \sim_{\text{lpo}} t_i\sigma$ for all $1 \leq i \leq n$ and therefore $s\sigma \succ_{\text{lpo}} t\sigma$ $\langle 1 \rangle$.

- If $s \succ_{\text{lpo}} t$ $\langle 2 \rangle$ then $f \succ g$ and $s \succ_{\text{lpo}} t_i$ for all $1 \leq i \leq n$. The induction hypothesis yields $s\sigma \succ_{\text{lpo}} t_i\sigma$ for all $1 \leq i \leq n$ and thus also $s\sigma \succ_{\text{lpo}} t\sigma$ by $\langle 2 \rangle$.

- If $s \succ_{\text{lpo}} t$ $\langle 3, i \rangle$ then $s_i \sim_{\text{lpo}} t$ or $s_i \succ_{\text{lpo}} t$. For the former $s_i\sigma \sim_{\text{lpo}} t\sigma$ holds by Lemma 5, for the latter we have $s_i\sigma \succ_{\text{lpo}} t\sigma$ by the induction hypothesis. So in both cases $s\sigma \succ_{\text{lpo}} t\sigma$ by $\langle 3, i \rangle$.

$\square$

**Lemma 8.** *$\succ_{\text{lpo}}$ is a proper order, i.e., it is irreflexive and transitive.*

*Proof.* Before proving transitivity and irreflexivity we note that whenever there are terms $s = f(s_1, \dots, s_m)$, $t = g(t_1, \dots, t_n)$ with $s \succ_{\text{lpo}} t$ then $s \succ_{\text{lpo}} t_i$ for all $1 \leq i \leq n$. The easy proof is the same as for strict precedences and can be found in [8]. Proving transitivity amounts to $s \succ_{\text{lpo}} t$ and $t \succ_{\text{lpo}} u$ implies $s \succ_{\text{lpo}} u$. Let $s = f(s_1, \dots, s_m)$ and $t = g(t_1, \dots, t_n) \succ_{\text{lpo}} u$. We show the result by induction on $||s|| + ||t|| + ||u||$. For the base case we have $||s|| = ||t|| = 1$, $||u|| = 0$. Clearly $t \succ_{\text{lpo}} u$ by $\langle 3, i \rangle$ and therefore the desired $s \succ_{\text{lpo}} t_i = u$ follows. For the inductive step assume $s' \succ_{\text{lpo}} t' \succ_{\text{lpo}} u'$ implies $s' \succ_{\text{lpo}} u'$ for all terms $s', t', u'$ with $||s'|| + ||t'|| + ||u'|| < k$ ($k > 2$) and $||s|| + ||t|| + ||u|| = k$.

- – Suppose $s \succ_{\text{lpo}} t \ \langle 1, i\rangle$ and $t \succ_{\text{lpo}} u \ \langle 1, j\rangle$. Then $u = h(u_1, \ldots, u_p)$ with $f \sim g \sim h$ and therefore $f \sim h$. Let $\mu = min\{i, j\}$. We show that $s \succ_{\text{lpo}} u \ \langle 1, \mu\rangle$ by proving (a) $s_l \sim_{\text{lpo}} u_l$ for all $1 \leq l < \mu$, (b) $s_\mu \succ_{\text{lpo}} u_\mu$, and (c) $s \succ_{\text{lpo}} u_l$ for all $\mu < l \leq p$.
    (a) Since $l < i, j$ clearly $s_l \sim_{\text{lpo}} u_l$.
    (b) The following three cases may appear: If $i = j = \mu$ then $s_\mu \succ_{\text{lpo}} t_\mu \succ_{\text{lpo}} u_\mu$ implies $s_\mu \succ_{\text{lpo}} u_\mu$ by the induction hypothesis. If $i = \mu < j$ or $j = \mu < i$ then $s_\mu \succ_{\text{lpo}} t_\mu \sim_{\text{lpo}} u_\mu$ or $s_\mu \sim_{\text{lpo}} t_\mu \succ_{\text{lpo}} u_\mu$ implies $s_\mu \succ_{\text{lpo}} u_\mu$ by Lemma 6.
    (c) Since $t \succ_{\text{lpo}} u_l$ the desired $s \succ_{\text{lpo}} u_l$ for all $i < l \leq n$ follows from the induction hypothesis.
  – Suppose $s \succ_{\text{lpo}} t \ \langle 1, i\rangle$ and $t \succ_{\text{lpo}} u \ \langle 1\rangle$, i.e., $s_j \sim_{\text{lpo}} t_j$ for all $1 \leq j < i$, $s_i \succ_{\text{lpo}} t_i$, $s \succ_{\text{lpo}} t_j$ for all $i < j \leq n$, $n > p$, and $t_j \sim_{\text{lpo}} u_j$ for all $1 \leq j \leq p$. If $i > p$ then clearly $m > p$. Transitivity of $\sim_{\text{lpo}}$ yields $s_j \sim_{\text{lpo}} u_j$ for all $1 \leq j \leq p$ and therefore $s \succ_{\text{lpo}} u \ \langle 1\rangle$. If $i \leq p$ then $s_j \sim_{\text{lpo}} u_j$ for all $1 \leq j \leq i$ by transitivity of $\sim_{\text{lpo}}$, $s_i \succ_{\text{lpo}} u_i$, and $s \succ_{\text{lpo}} u_j$ for all $i < j \leq p$ by the induction hypothesis and consequently $s \succ_{\text{lpo}} u \ \langle 1, i\rangle$.
  – The case where $s \succ_{\text{lpo}} t \ \langle 1\rangle$ and $t \succ_{\text{lpo}} u \ \langle 1, i\rangle$ is similar to the one above.
  – Suppose $s \succ_{\text{lpo}} t \ \langle 1\rangle$ and $t \succ_{\text{lpo}} u \ \langle 1\rangle$. Then $m > n$, $s_i \sim_{\text{lpo}} t_i$ for all $1 \leq i \leq n$, $n > p$, and $t_i \sim_{\text{lpo}} u_i$ for all $1 \leq i \leq p$. Transitivity of $\sim_{\text{lpo}}$ yields $s_i \sim_{\text{lpo}} u_i$ for all $1 \leq i \leq p$ and together with $m > n > p$ proves $s \succ_{\text{lpo}} u \ \langle 1\rangle$.
- Suppose $s \succ_{\text{lpo}} t \ \langle 1\rangle$ and $t \succ_{\text{lpo}} u \ \langle 2\rangle$. We have $u = h(u_1, \ldots, u_p)$ with $f \sim g \succ h$ and $t \succ_{\text{lpo}} u_i$ for all $1 \leq i \leq p$. The induction hypothesis yields $s \succ_{\text{lpo}} u_i$ for all $1 \leq i \leq p$ and because $f \succ h$ also $s \succ_{\text{lpo}} u$ holds by $\langle 2\rangle$.
- If $s \succ_{\text{lpo}} t \ \langle 2\rangle$ and $t \succ_{\text{lpo}} u \ \langle 1\rangle$ or $\langle 2\rangle$ then $f \succ g$ and $u = h(u_1, \ldots, u_p)$ with $g \succsim h$. As $f \succ h$ and $s \succ_{\text{lpo}} u_i$ for all $1 \leq i \leq p$ by the induction hypothesis also $s \succ_{\text{lpo}} u$ holds by $\langle 2\rangle$.
- If $s \succ_{\text{lpo}} t$ and $t \succ_{\text{lpo}} u \ \langle 3\rangle$ then $s \succ_{\text{lpo}} t_i \succsim_{\text{lpo}} u$ and thus $s \succ_{\text{lpo}} u$ either by the induction hypothesis or Lemma 6. If $s \succ_{\text{lpo}} t \ \langle 3\rangle$ and $t \succ_{\text{lpo}} u$ then $s_i \succsim_{\text{lpo}} t \succ_{\text{lpo}} u$ and again $s \succ_{\text{lpo}} u$ either by the induction hypothesis or Lemma 6.

$\succ_{\text{lpo}}$ is irreflexive: By induction on $||t||$ we show that $t \succ_{\text{lpo}} t$ does not hold. In the base case $||t|| = 0$ and thus $t = x$ for some $x \in \mathcal{V}$. Therefore $t \succ_{\text{lpo}} t$ cannot hold. For the inductive step assume that $t' \succ_{\text{lpo}} t'$ does not hold for terms $t'$ with $||t'|| < k$ ($k > 0$) and let $||t|| = k$. For a proof by contradiction assume that $t \succ_{\text{lpo}} t$ holds. If $t \succ_{\text{lpo}} t \ \langle 1, i\rangle$ then $t_i \succ_{\text{lpo}} t_i$ contradicting the induction hypothesis. Also $t \succ_{\text{lpo}} t \ \langle 1\rangle$ leads to a contradiction because both terms have equally many arguments. If $t \succ_{\text{lpo}} t$ $\langle 2 \rangle$ then $f \succ f$ contradicts the irreflexivity of $\succ$. Finally, if $t \succ_{\text{lpo}} t$ $\langle 3 \rangle$ then $t_i \gtrsim_{\text{lpo}} t$. Because $t \succ_{\text{lpo}} t_i$ $\langle 3 \rangle$, $t_i \succ_{\text{lpo}} t_i$ follows either by Lemma 6 or transitivity of $\succ_{\text{lpo}}$ and contradicts the induction hypothesis. $\square$

**Lemma 9.** $\succ_{\text{lpo}}$ *has the subterm property.*

*Proof.* As $\rhd_{\text{emb}}$ has the subterm property it suffices to show that $\rhd_{\text{emb}} \subseteq \succ_{\text{lpo}}$. If $l \to r \in \mathcal{E}mb$ then $r$ is an argument of $l$ and thus $l \succ_{\text{lpo}} r$ $\langle 3 \rangle$. Since $\succ_{\text{lpo}}$ is a rewrite order $\rhd_{\text{emb}} \subseteq \succ_{\text{lpo}}$. $\square$

**Corollary 1.** $\succ_{\text{lpo}}$ *is a simplification order.*

# B Implementation Details

The software is mainly written in the functional programming language `OCaml` and integrates some already existing modules for parsing, terms, graph theory, etc. from T$\top$T [10] which has been developed by Nao Hirokawa and Aart Middeldorp. The non `OCaml` part is the state of the art SAT solver MiniSat [3] which has been employed to test satisfiability of the constraint formula $B(X, Y)$.

## B.1 The DIMACS Input Format

Although the general input format for MiniSat is rather simple a short description is provided here because many references on the web are too imprecise and directly lead to some pitfalls.

**Example 6.** Consider the propositional CNF formula $p \wedge (\neg q \vee r) \wedge (q \vee \neg r) \wedge q$. The variables are represented by integer values. E.g., $p$ by 1, $q$ by 2, and $r$ by 3. Negated variables are encoded by the corresponding negative value, e.g., $\neg q$ by $-2$ and so on. Typically a file starts with some comments and these lines begin with `c`. After that a line like `p cnf 3 4` indicates the format of the input file. Up to version 1.13 MiniSat also supported the more general `sat` format but the most recent version demands a CNF input (therefore keyword `cnf`). The specified number `3` indicates the number of variables (an upper bound) and `4` reflects the number of conjuncts. At least for the latest version of MiniSat (v1.14) these two numbers are optional. After that line conjuncts in their integer encoding follow. Important is that every conjunct is trailed by a 0 whereas the newline is optional.

```
c This is a comment line
c
p cnf 3 4
1 0
-2 3 0
2 -3 0
2 0
```

Listing 1: A simple SAT instance in the DIMACS format.

## B.2 Interfacing MiniSat with OCaml

Chapter 18 of the OCaml reference manual [7] only explains how to interface C code with OCaml. SWIG [9] claims that it is able to interface C++ with OCaml but not even the examples given in the documentation could be compiled correctly. Anyway, if no complex data structures are shared the following explanations are totally sufficient for a working interface. The only differences compared to [7] is the keyword `extern "C"` in the declaration of the functions to prevent the C++ name mangling mechanism and the usage of the `g++` compiler instead of `cc`. In the sequel we first describe how to build a shared C++ library. After that a sample OCaml program gets linked against that library. We start with the necessary C++ files. As already mentioned it is important to decorate the declarations of the functions which are intended to be called from OCaml code with `extern "C"`. The two listings below present C++ code for the declaration (`test.h`) as well as the implementation (`test.c`) of two functions.

```
extern "C" void test()              ;
extern "C" int  add ( int a, int b);
```

Listing 2: test.h

```
#include"test.h"
#include<iostream>

using namespace std;

void test()
  {
  cout << "it c++ works" << endl;
  }//end test()

int add( int a, int b)
  {
  return a + b;
  }//end add()
```

Listing 3: test.c

These two files are linked into a dynamic shared library with the two commands

```
$ g++ -fPIC -g -c -Wall test.c
$ g++ -shared -o dllmylib.so test.o
```

where `-fPIC` is the flag for 'position independent code', `-g` enables debugging information, and `-Wall` tells the compiler to show all warnings. The second command then produces the shared library (flag `-shared`) named `dllmylib.so` from the object file `test.o`. So producing the C++ library was easy. Let's now face the interface for the functions which consists of an interface file (`interface.mli`) which declares the functions and their corresponding types and the so called stub (`interface_stubs.c`). Concerning the interface file the keyword `external` tells the `OCaml` compiler that these functions are implemented elsewhere. After that keyword the name of the function in the `OCaml` file is specified together with its type. After the `=` the name of the corresponding C function is needed. As one can see from the example the function names in the two implementations might differ but need not.

```
external otest: unit -> unit        = "test"
external add   : int   -> int -> int = "add"
```

Listing 4: interface.mli

The stub has to include `test.h` as well as other `OCaml` specific header files.

```
#include "test.h"
#include <caml/mlvalues.h>
#include <caml/memory.h>
#include <caml/alloc.h>
#include <caml/custom.h>

value caml_otest(value unit)
{
  CAMLparam1 (unit);
  test();
  CAMLreturn (Val_unit);
}

value caml_add(value a, value b)
{
  CAMLparam2 (a, b);
  CAMLreturn (Val_int(add(Int_val(a), Int_val(b))));;
}
```

Listing 5: interface_stubs.c

These two files should be compiled with:

```
$ ocamlc interface.mli
$ g++ -c -I/usr/local/lib/ocaml interface_stubs.c
```

Note that the include path of `OCaml` may vary. With the following main file

```
open Interface;;
let main () =
  otest ();
  let (a,b) = (3,4) in
  let n = add a b in
  Printf.printf "Sum of %d and %d equals %d\n" a b n;
;;

main();;
```

Listing 6: main.ml

an executable can be compiled with the command

```
$ ocamlc -o a.out main.ml -dllib -lmylib
```

Executing `a.out` should then produce an output similar to

```
it c++ works



Sum of 3 and 4 equals 8
```

Listing 7: Example output.

The attentive reader may have noticed that the sum of 3 and 4 actually equals 7 and not 8. In the interface this error occurs because `OCaml` and C++ have different representations of integers. As the interface above was constructed according to the description in [7] the problem probably is not due to this interface but the `OCaml` implementation. When an integer from `OCaml` code is passed to C++ code one bit is added (namely a 1 at the position of the least significant bit). Therefore not 3 and 4 are added but 7 and 9. From the result 16 the least significant bit is chopped off again and therefore the result 8 is reported.

Because memory allocation for strings also did not work without troubles all data sharing is done by Unix pipes. The file descriptors are redirected to stdout and stdin respectively. As a consequence of that up to 30% of the execution time is needed to hand over the CNF formula to MiniSat. By a thorough interface — which might be some work — that bottleneck should be removable.